\documentstyle[12pt]{article}
\def\be{\begin{equation}}
\def\ee{\end{equation}}
\def\bea{\begin{eqnarray}}
\def\eea{\end{eqnarray}}
\def\beqa{\begin{eqnarray}}
\def\eeqa{\end{eqnarray}}
\def\beq{\begin{equation}}
\def\eeq{\end{equation}}
\def\beqal{\begin{eqnarray}\label}
\def\beql{\begin{equation}\label}

\def\R{\mbox{\rm I\kern-.18em R}}

\def\P{\mbox{\rm I\kern-.18em P}}
\def\uno{\mbox{1 \kern-.59em {\rm l}}}
\def\Ds{\ {\big / \kern-.70em D}}
\def\ds{\big / \kern-.90em {\ \p} }

\def\cDs{\ {\big / \kern-.70em {\cal D}}}
\def\Z{{Z \kern-.45em Z}}
\def\Q{{\kern .1em {\raise .47ex \hbox{$\scriptscriptstyle |$}}
\kern -.35em {\rm Q}}}

\def\Tr{\mbox{\rm Tr}}

\def\p{\partial}

\def\om{\omega}

\def\1{\dot{1}}
\def\2{\dot{2}}

\font\mybb=msbm10 at 12pt

\def\bb#1{\hbox{\mybb#1}}
\def\Z {\bb{Z}}
\def\C {\bb{C}}

\begin{document}\begin{titlepage}

\hfill{DFPD00/TH/45}

\hfill{hep-th/0103246}

\vspace{1cm}
\begin{center}
{\large \bf MULTI-INSTANTON MEASURE FROM RECURSION RELATIONS}

\vspace{.3cm}

{\large\bf IN N=2 SUPERSYMMETRIC YANG-MILLS THEORY}

\end{center}
\vspace{1.5cm}
\centerline{MARCO MATONE}
\vspace{0.8cm}
\centerline{\it Department of Physics ``G. Galilei'' - Istituto Nazionale di
Fisica Nucleare}
\centerline{\it University of Padova}
\centerline{\it Via Marzolo, 8 - 35131 Padova, Italy}
\centerline{matone@pd.infn.it}

\vspace{2cm}
\centerline{\sc ABSTRACT}

\vspace{0.6cm} \noindent By using the recursion relations found in the
framework of $N=2$ Super Yang-Mills theory with gauge group $SU(2)$, 
we reconstruct the structure of the instanton moduli space and its 
volume form for all winding numbers. The construction is reminiscent 
of the Deligne-Knudsen-Mumford compactification and uses an analogue 
of the Wolpert restriction phenomenon which arises in the case of 
moduli spaces of Riemann surfaces.
\end{titlepage}
\newpage
\setcounter{footnote}{0}
\renewcommand{\thefootnote}{\arabic{footnote}}
\noindent In \cite{sw} the entire nonperturbative contribution to the
holomorphic part of the Wilsonian effective action was computed for
$N=2$ supersymmetric theories with gauge group $SU(2)$, using 
ans\"atze dictated by physical intuitions. There are several aspects 
of the Seiberg-Witten (SW) model \cite{sw} which are related to the 
theory of moduli spaces of Riemann surfaces. In particular, here, we 
will consider the recursion relations for nonperturbative 
(instanton) contributions to the $N=2$ Super Yang-Mills (SYM) 
effective prepotential \cite{alfa} and will compare them with the 
recursion relations for the Weil-Petersson (WP) volumes of punctured 
Riemann spheres. In the SW model there exists a relation
 the modulus $u=\langle\Tr\phi^2\rangle$ and
the effective prepotential \cite{alfa} (see also \cite{vari}). This 
allowed to prove the SW conjecture by using the reflection symmetry 
of vacua \cite{BMT}. On the other hand, it is rather surprising
that, while on one side all the instanton coefficients have been
computed in \cite{alfa}, explicit calculations have been performed
only in the one and two-instanton background
\cite{fp,DKM,ft}, while the above mentioned relation has been shown
to hold to all instanton orders \cite{DKMlett,oveveste}. The problem 
for instanton number $k\geq 3$ seems extremely difficult to solve. 
Indeed, the ADHM constraint equations become nonlinear and have not 
been explicitly solved up to now. Moreover, neither the structure of 
the moduli space, nor the volume form are known. The instanton 
measure for all winding numbers has been written in \cite{dkmdelta}, 
but only in an implicit form ({\it i.e.} by implementing the bosonic 
and fermionic ADHM constraints through the use of Dirac delta 
functions), which in some special cases allows to extract 
information on the instanton moduli space
\cite{DoreyAdS}. However, the mathematical challenging problem of
finding the explicit structure of the instanton moduli space for 
generic winding numbers still remains unsolved. On the other hand, 
the simple way in which the recursion relations have been derived, 
strongly suggests that there may be some mechanism which should make 
the explicit calculations possible. The investigation of such 
mechanism would provide important information on the structure of 
the instanton moduli space (of which only the boundary \`a la 
Donaldson-Uhlenbeck is known for generic winding number 
\cite{Freed,dk,maciocia}) and of the associated volume form. In 
particular, even if the integrals seem impossible to compute, 
(actually, as we stated before we know neither the structure of the 
space nor the volume form), the existence of recursion relations and 
the simple way in which they arise, seem to suggest that these 
integrals could be easy to compute because of some underlying 
geometrical recursive structure. It has been claimed for some time, 
but only recently proven \cite{bftt}, that the nonperturbative 
contributions to $u$ actually can be written as total derivatives, 
{\it i.e.} as pure boundary terms, on the moduli space. If the 
boundary is composed by moduli spaces of instantons of lower winding 
number times zero-size instantons moduli spaces, as it happens in 
the Donaldson-Uhlenbeck compactification, this would immediately 
provide, in the case of a suitable volume form, a recursion relation.

\noindent
We will now see how the similar problem one finds in computing the
WP volumes of punctured spheres has been solved thanks to the 
recursive structure of the Deligne-Knudsen-Mumford (DKM) boundary 
and to the peculiar nature of the WP 2-form. The main analogy we 
will display, concerns the volume of moduli space of
$n$-punctured Riemann spheres $\Sigma_{0,n}=\widehat
{\C}\backslash\{z_1,\ldots,z_n\}$, $n\ge 3$, where $\widehat
{\C}\equiv {\C}\cup\{\infty\}$. Their moduli space is the space of
classes of isomorphic $\Sigma_{0,n}$'s, that is
\begin{equation}
{\cal M}_{0,n}= \{(z_1,\ldots,z_{n})\in \widehat{\C}^{n}|z_j\ne
z_k\; {\rm for}\; j\ne k\}/Symm(n)\times PSL(2,{\C})\ \ ,
\label{modulisp}\end{equation} where ${Symm}(n)$ acts by permuting
$\{z_1,\ldots,z_n\}$ whereas $PSL(2,{\C})$ acts as a linear
fractional transformation. Using $PSL(2,\C)$ symmetry we can recover
the ``standard normalization'': $z_{n-2}=0$, $z_{n-1}=1$ and
$z_{n}=\infty $. The classical Liouville tensor or Fuchsian
projective connection is
\beq
T^F(z)=\left\{J_H^{-1},z\right\}=\varphi_{{cl}\,zz}-{1\over 2}
\varphi_{{cl}\,z}^2\ \ .
\eeq
In the case of the punctured Riemann sphere we have
\beq T^F(z)=\sum_{k=1}^{n-1}\left({1\over 2(z-z_k)^2}+ {c_k\over
z-z_k}\right)\ \ ,
\eeq where the coefficients $c_1,\ldots c_{n-1}$, called {\it
accessory parameters}, satisfy the constraints
\beq
\sum_{j=1}^{n-1}c_{j}=0\ \ ,\qquad
\sum_{j=1}^{n-1}z_jc_j=1-{n\over 2}\ \ . \eeq
These parameters are
defined on the space
\begin{equation}
 V^{(n)}=\{(z_1,\ldots,z_{n-3})\in
{\C}^{n-3}|z_j\ne 0,1\ \ ; z_j\ne z_k\ \ ,\; {\rm for}\; j\ne k\}\ \
.
\label{star}\end{equation} Note that
\begin{equation}
{\cal M}_{0,n}\cong V^{(n)}/{Symm}(n)\ \ , \label{mdls}\end{equation}
where the action of $Symm(n)$ on $V^{(n)}$ is defined by comparing
(\ref{modulisp}) with (\ref{mdls}).

\noindent
Let us now consider the compactification $\overline V^{(n)}$ \`{a} 
la DKM \cite{DeligneKnudsenMumford}\cite{Z}. The divisor at the 
boundary
\beq
D=\overline V^{(n)}\backslash V^{(n)}\ \ , \eeq decomposes in the
sum of divisors $D_1$,\ldots,$D_{[n/2]-1}$, which are subvarieties 
of real dimension $2n-8$. The locus $D_k$ consists of surfaces that 
split, upon removal of the node, into two Riemann spheres with
$k+2$ and $n-k$ punctures. In particular, $D_k$ consists of $C(k)$
copies of the space $\overline V^{(k+2)}\times \overline V^{(n-k)}$ 
where $C(k)= \pmatrix{ n\cr k+1}$, for $k=1,\ldots,{(n-3)/2}$,
$n$ odd. In the case of even $n$ the unique difference is for
$k=n/2-1$, for which we have $C(n/2-1)={1\over 2}\pmatrix{n\cr
n/2}$. An important property of the divisors $D_k$'s is that their 
image provides a basis in $H_{2n-8}(\overline{\cal M}_{0,n},{\R})$. 
The WP volume is
\beq {\rm Vol}_{WP}\left({\cal M}_{0,n}\right)={1\over (n-3)!} \int_{
\overline{\cal M}_{0,n}}{\omega_{WP}^{(n)}}^{n-3}= {1\over (n-3)!}
\left[\omega_{WP}^{(n)}\right]^{n-3}\cap \left[\overline{\cal
M}_{0,n}\right]\ \ , \eeq where $\cap$ is topological cup product. 
It has been shown that \cite{Z}
\beq
{\rm Vol}_{WP}\left({\cal M}_{0,n}\right)={1\over n!} {\rm
Vol}_{WP}\left(V^{(n)}\right)={\pi^{2(n-3)}
 V_n\over n!(n-3)!}\ \ ,\qquad n\ge 4\ \ ,
\eeq where $V_n=\pi^{2(3-n)}\left[\omega_{WP}^{(n)}\right]^{n-3}\cap
\left[\overline{V}^{(n)}\right]$ satisfies the recursion relations
\begin{equation}  V_3=1\ \ ,\qquad
V_n={1\over 2}\sum_{k=1}^{n-3}{k(n-k-2) \over n-1}
\left(\begin{array}{c} n\\ k+1 \end{array}\right)
\left(\begin{array}{c} n-4\\ k-1 \end{array}\right)
 V_{k+2}V_{n-k}\ \ ,\qquad
 n \ge 4.\label{51}\end{equation}
These relations are a consequence of two basic properties. The first 
one is that the boundary of $\overline{\cal M}_{0,n}$ in the DKM 
compactification is the union of product of moduli spaces of lower 
order. The second one is the restriction phenomenon satisfied by the 
WP 2-form. A property discovered by Wolpert in
\cite{wolpertis} (see also the Appendix of \cite{ma}).
The basic idea is to start with the natural embedding
\beq
i: \overline{V}^{(m)}\to\overline{V}^{(m)}\times
* \to \overline{V}^{(m)}\times \overline{V}^{(n-m+2)}
\to \partial \overline{V}^{(n)} \to\overline{V}^{(n)}\ \ ,
\qquad n>m\ \ ,
\eeq
where $*$ is an arbitrary point in $\overline{V}^{(n-m+2)}$, it 
follows that \cite{wolpertis}
\begin{equation}
\left[\omega_{WP}^{(m)}\right]= i^*\left[\omega_{WP}^{(n)}
\right]\ \ , \qquad n>m\ \ .
\label{gdtdte}\end{equation}
There is a similarity between the above recursion relations for the 
WP volumes and the recursion relations satisfied by the instanton 
coefficients. To see this let us recall that, in the case of the WP 
volumes, it has been derived in \cite{ma} a nonlinear ODE satisfied 
by the generating function for the WP volumes
\beq
g(x)=\sum_{k=3}^\infty a_k x^{k-1}\ \ , \label{geneterg}
\eeq
where
\begin{equation}
 a_k= {V_k\over (k-1)((k-3)!)^2}\ \ ,\qquad k\ge 3\ \ ,
\label{rnm2}\end{equation} so that (\ref{51}) assumes the simple
form
\begin{equation}
a_3=1/2\ \ ,\qquad a_n={1\over 2}{n(n-2)\over (n-1)(n-3)}
\sum_{k=1}^{n-3}a_{k+2}a_{n-k}\ \ ,\qquad n\ge
4\ \ .\label{51al}\end{equation} These recursion relations have been 
the starting point to formulate a nonperturbative model of Liouville 
quantum gravity as Liouville F-model \cite{ma} (see also 
\cite{bomama}). In particular, this formulation has been obtained as 
a deformation of WP volumes. Furthermore, it has been conjectured 
that the relevant integrations on the moduli space of higher genus 
Riemann surfaces reduce to integrations on the moduli space of 
punctured Riemann spheres (see also 
\cite{manin}\cite{Givental}\cite{underground}). One can check that  
Eq.(\ref{51al}) implies that the function $g$ satisfies the ODE 
\cite{ma}
\begin{equation}
x(x-g)g''=xg^{'2}+(x-g)g'\ \ .
\label{51a}\end{equation}
Remarkably, it has been shown in \cite{manin} that this nonlinear 
ODE is essentially the inverse of a linear one. More
 precisely, defining $g=x^2\partial_x x^{-1}
h$, one has that (\ref{51a}) implies \beq xh''-h'= (xh'-h)h''\ \ . 
\label{lalineare}\eeq Differentiating (\ref{lalineare}) we get
\beq yy''=xy^3\ \ ,\label{onan} \eeq where
$y=h'$. Then, interchanging the r\^{o}les of $x$ and $y$, (\ref{onan})
transforms into the Bessel equation
\beq y\frac{d^2 x}{d y^2}+x=0\ \ .
\label{bessel} \eeq
It has been suggested in \cite{manin} that the appearance of such a 
linear ODE may be related to the ``mirror phenomenon". The above 
structure is reminiscent of the above derived in SW theory. In 
particular, in the case of WP volumes one starts evaluating the 
recursion relations by means of the DKM compactification and the 
Wolpert restriction phenomenon \cite{Z}, then derives the associated 
nonlinear ODE \cite{ma} and end to a linear ODE \cite{manin} which 
is obtained by essentially inverting it. In the SW model, one starts 
by observing that the $a^D(u)$ and 
$a(u)$ moduli satisfy a linear ODE \cite{KLT}, inverts it to obtain
a nonlinear one satisfied by $u(a)$ and then finds recursion 
relations for the coefficients of the expansion of $u(a)$ 
\cite{alfa}. The final point stems from the observation \cite{alfa} 
that $u$ and ${\cal F}$ are related in a simple way which allows one 
to consider the derived recursion relation as a relation for the 
instanton contributions to the prepotential ${\cal F}$. The above 
similarity suggests to reconstruct the instanton moduli space and 
its measure starting from the recursion relations \cite{alfa}
\beq
{\cal G}_{n+1}={1\over 8{\cal G}_0^2(n+1)^2}
\left[(2n-1)(4n-1){\cal G}_n+2{\cal G}_0
\sum_{k=0}^{n-1}c_{k,n}{\cal G}_{n-k}{\cal G}_{k+1}
-2\sum_{j=0}^{n-1}\sum_{k=0}^{j+1}d_{j,k,n}{\cal G}_{n-j} {\cal 
G}_{j+1-k}{\cal G}_{k}\right]\ \ ,
\label{recursion2}
\eeq
where $n\geq 0$, ${\cal G}_0=1/2$ and
\beq
c_{k,n}=2k(n-k-1)+n-1\ \ ,
\qquad
d_{j,k,n}= [2(n-j)-1][2n-3j-1+2k(j-k+1)]\ \ .
\eeq
It is still possible to rewrite some apparently cubic terms in the 
third term on the r.h.s. as quadratic ones and absorb them in the 
second term on the r.h.s. of (\ref{recursion2}), obtaining thus
\beq
{\cal G}_{n+1}={1\over 2(n+1)^2}
\left[
(2n-1)(4n-1){\cal G}_n +\sum_{k=0}^{n-1}b_{k,n}{\cal G}_{n-k}{\cal
G}_{k+1} -2\sum_{j=1}^{n-1}\sum_{k=1}^{j}d_{j,k,n}{\cal G}_{n-j} 
{\cal G}_{j+1-k}{\cal G}_{k}\right]\ \ ,
\label{recursion3}
\eeq
where $b_{k,n}=c_{k,n}-2d_{k,0,n}$ and we have exploited the fact
that $d_{k,0,n}=d_{k,k+1,n}$. Let us now consider the volume ${\cal
G}_{n}$ of the moduli space of an instanton configuration of winding
number $n$. We now start showing that ${\cal G}_{n}$ can be
expressed as
\beq\label{uno}
{\cal
G}_{n}=\int_{\overline{V}_I^{(n)}}\bigwedge_{k=1}^{X(n)}\om_I^{(n)}=
[\om_I^{(n)}]^{X(n)}\cap[\overline{V}_I^{(n)}]\ \ ,
\eeq
where $\om_I^{(n)}$ is a 2-form defined on the
$n$-instanton moduli space and $\overline{V}_I^{(n)}$ is a suitable
compactification of $V_I^{(n)}$, which, together its complex 
dimension $X(n)$, will be fixed later. Let
${\cal D}^{(n+1)}_{\om}$ be the $[2X(n+1)-2]$-cycle Poincar\'e dual to the 
``instanton'' class $[\om_I^{(n+1)}]$. That is, 
$[\om_I^{(n+1)}]=c_1([{\cal D}^{(n+1)}_{\om}])$ where, as usual,
$[{\cal D}]$ denotes the line bundle associated to a given divisor
${\cal D}$ (see, for example, \cite{GriffithsHarris})
and $c_1$ denotes the first Chern class.
By Poincar\'e duality it is possible to recast 
(\ref{uno}) in the form
\beqa\label{duee}
{\cal G}_{n+1}&=&[\om_I^{(n+1)}]^{X(n+1)-1}\cap([\om_I^{(n+1)}]
\cap[\overline{V}_I^{(n+1)}])
=[\om_I^{(n+1)}]^{X(n+1)-1}\cap[{\cal
D}^{(n+1)}_{\om}\cdot\overline{V}_I^{(n+1)}]
\nonumber\\
&=&[\om_I^{(n+1)}]^{X(n+1)-1}\cap[{\cal D}^{(n+1)}_{\om}]\ \ ,
\eeqa
where $\cdot$ denotes the topological intersection. In order to make 
contact with the recursion relation for the ${\cal G}_n$'s, we
define the following compactification
\beq
{\cal D}^{(n+1)}=\overline{V}_I^{(n+1)}/V_I^{(n+1)}=
\sum_{j=0}^{n-1}{\cal D}_{1,j}+\sum_{j=1}^{n-1}\sum_{k=1}^{j}
{\cal D}_{2,j,k}+{\cal D}_{3,n} \ \  ,
\label{bhos}\eeq
in the sense of cycles on orbifolds, where
\beqa
{\cal D}_{1,j}&=&c^{(1)}_{n,j}\overline{V}_I^{(n-j)}\times
\overline{V}_I^{(j+1)}\ \ ,\nonumber\\
{\cal D}_{2,j,k}&=&c^{(2)}_{n,j,k}
\overline{V}_I^{(n-j)}\times\overline{V}_I^{(j+1-k)}\times
\overline{V}_I^{(k)}\times
\overline{V}_I^{(1)}
\ \ ,\nonumber\\
{\cal D}_{3,n}&=&c^{(3)}_{n}\overline{V}_I^{(n)}\times
\overline{ V}_I^{(1)}\ \ .
\eeqa
Observe that in the above decomposition the products of subvarieties 
generally appear twice. Furthermore, note that we used ${\cal 
D}_{3,n}$ to simplify the calculations, however it can be included 
either in ${\cal D}_{1,0}$ or ${\cal D}_{1,n-1}$ by changing the 
coefficients. Let us now expand 
${\cal D}^{(n+1)}_{\om}$ in terms of the divisors at the boundary of 
the moduli space, namely
\beq\label{boh}
{\cal D}^{(n+1)}_{\om}=\sum_{j=0}^{n-1}d^{(1)}_{n,j}{\cal D}_{1,j}+
\sum_{j=1}^{n-1}\sum_{k=1}^{j}d^{(2)}_{n,j,k}{\cal D}_{2,j,k}+
d^{(3)}_{n}{\cal D}_{3,n}\ \ .
\eeq
One can see that a check on the outlined procedure uniquely fixes 
$X(n)$ to be
\beq
X(n)=2n-1\ \ .
\eeq
By (\ref{duee}) we have
\beq\label{quellaaggiuntaduee}
{\cal G}_{n+1}=
\sum_{j=0}^{n-1}d^{(1)}_{n,j}[\om_I^{(n+1)}]^{2n}\cap[{\cal D}_{1,j}]+
\sum_{j=1}^{n-1}\sum_{k=1}^{j}d^{(2)}_{n,j,k}
[\om_I^{(n+1)}]^{2n}\cap[{\cal D}_{2,j,k}]+ 
d^{(3)}_{n}[\om_I^{(n+1)}]^{2n}\cap[{\cal D}_{3,n}]\ \ .
\eeq
Let us consider the following natural embedding
\beq
i: \overline{V}_I^{(m)}\to
\overline{V}_I^{(m)}\times * \to \overline{V}_I^{(m)}\times
\overline{V}_I^{(n-m)} \to \partial \overline{V}_I^{(n)} \to
\overline{V}_I^{(n)}\ \ , \qquad n>m\ \ ,
\label{gdtdteunosse}
\eeq
where $*$ is an arbitrary point in $\overline{V}_I^{(n-m)}$. We now
impose the following constraint
\beq
\left[\omega_{I}^{(m)}\right]= i^*\left[\omega_I^{(n)}\right]\ \ ,
\qquad n>m\ \ .
\label{gdtdtebisse}
\eeq
Let us elaborate the three terms on the r.h.s. of 
(\ref{quellaaggiuntaduee}). By (\ref{gdtdteunosse}) and 
(\ref{gdtdtebisse}) the general contribution in the first term reads
\beqa
&&[\om_I^{(n+1)}]^{2n}\cap[\overline{V}_I^{(n-j)}\times
\overline{V}_I^{(j+1)}]=[\om_I^{(n-j)}+\om_I^{(j+1)}]^{2n}\cap
[\overline{V}_I^{(n-j)}\times\overline{V}_I^{(j+1)}]\nonumber\\
&&=C_{j,n}([\om_I^{(n-j)}]^{2(n-j)-1}\cap 
[\overline{V}_I^{(n-j)}])([\om_I^{(j+1)}]^{2j+1}\cap 
[\overline{V}_I^{(j+1)}])=C_{j,n}{\cal G}_{n-j}{\cal G}_{j+1}
\ \ ,
\eeqa
where $C_{j,n}=\pmatrix{2n\cr 2(n-j)-1}$. The second term has the 
form
\beqa
&&[\om_I^{(n+1)}]^{2n}\cap[\overline{V}_I^{(n-j)}\times
\overline{V}_I^{(j+1-k)}\times\overline{V}_I^{(k)}
\times\overline{ V}_I^{(1)}]\nonumber\\
&&=[\om_I^{(n-j)}+\om_I^{(j+1-k)}+\om_I^{(k)}+
\om_I^{(1)}]^{2n}\cap
[\overline{V}_I^{(n-j)}\times
\overline{V}_I^{(j+1-k)}\times\overline{ V}_I^{(k)}
\times\overline{V}_I^{(1)}]\nonumber\\
&&=D_{j,n,k} ([\om_I^{(n-j)}]^{2(n-j)-1}\cap[\overline{V}_I^{(n-j)}])
([\om_I^{(j+1-k)}]^{2(j-k)+1}\cap
[\overline{V}_I^{(j+1-k)}])([\om_I^{(k)}]^{2k}
\cap[\overline{V}_I^{(k)}])([\om_I^{(1)}]
\cap[\overline{V}_I^{(1)}])\nonumber\\
&&={D_{j,n,k}\over 4}{\cal G}_{n-j}{\cal G}_{j+1-k}{\cal G}_{k}
\ \ ,
\eeqa
where $D_{j,n,k}= 2k\pmatrix{2n\cr 2k}\pmatrix{2n-2k\cr 2(n-j)-1}$
and we used the fact that ${\cal G}_1=1/4$. Finally, the last term is
\beq
[\om_I^{(n+1)}]^{2n}\cap[\overline{V}_I^{(n)}\times
\overline{V}_I^{(1)}]={n\over 2}{\cal G}_{n}\ \ .
\eeq
In this way we can recast the recursion relations as
\beqa
{\cal G}_{n+1}&=&\sum_{k=0}^{n-1}
\pmatrix{2n\cr 2(n-k)-1}d^{(1)}_{n,k}c^{(1)}_{n,k}
{\cal G}_{n-k}{\cal G}_{k+1}+
\sum_{j=1}^{n-1}\sum_{k=1}^{j}{k\over 2}
\pmatrix{2n\cr 2k}\cdot\nonumber\\
&&\cdot\pmatrix{2(n-k)\cr 2(n-j)-1} 
d^{(2)}_{n,j,k}c^{(2)}_{n,j,k}{\cal G}_{n-j}{\cal G}_{j+1-k}{\cal 
G}_k+{n\over 2}d^{(3)}_{n}c^{(3)}_{n}{\cal G}_n\ \ ,
\eeqa
where now $n\geq 1$ (and ${\cal G}_1=1/4$). Comparing with 
(\ref{recursion3}) we have
\beqa
&&d^{(1)}_{n,k}c^{(1)}_{n,k}\pmatrix{2n\cr 2(n-k)-1}=
\frac{b_{k,n}}{2(n+1)^2}
\ \ ,\nonumber\\
&&d^{(2)}_{n,j,k}c^{(2)}_{n,j,k}\pmatrix{2n\cr 2k}
\pmatrix{2(n-k)\cr 2(n-j)-1}=-\frac{2d_{j,k,n}}{k(n+1)^2}\ \ ,
\nonumber\\
&&d^{(3)}_{n}c^{(3)}_{n}=\frac{(2n-1)(4n-1)}{n(n+1)^2}\ \ .
\eeqa
\vspace{.5cm}

\noindent
{\bf Acknowledgements.} We would like to thank D. Bellisai for many 
fruitful discussions. Work partly supported by the European 
Commission TMR programme ERBFMRX--CT96--0045.

\end{document}